\documentstyle[aps,prl,twocolumn,epsfig]{revtex}

\begin{document}

\draft

\title{Fidelity balance in quantum operations}
\author{Konrad Banaszek}
\address{Center~for~Quantum~Information and
Rochester~Theory~Center for Optical Science and Engineering,\\
University~of~Rochester, Rochester NY 14627\\
and Instytut Fizyki Teoretycznej, Uniwersytet Warszawski, Ho\.{z}a 69,
PL--00--681 Warszawa, Poland}
\date{\today}

\maketitle

\begin{abstract}
I derive a tight bound between the quality
of estimating the state of a single copy of a $d$-level system, and
the degree the initial
state has to be altered in course of this procedure.
This result provides a complete analytical description
of the quantum mechanical trade-off between the
information gain and the quantum state disturbance 
expressed in terms of mean fidelities.
I also discuss consequences of this bound for quantum teleportation
using nonmaximally entangled states.
\end{abstract}

As a general rule, the more information is obtained from an operation
on a quantum system, the more its state has to be altered. This heuristic
statement was first exemplified by the Heisenberg microscope
{\em gedankenexperiment} \cite{Heisenberg},
where the spatial resolution of the apparatus
was shown to scale inversely with the uncertainty of the momentum
transfered during the observation. Presently, the disturbance caused
by the information gain has become an issue of practical significance,
as it underlies the security of quantum key distribution 
\cite{Cryptography}.

The balance between the information gain and the state disturbance
attracts currently a lot of interest,
particularly in the context of quantum cryptography
\cite{Balance}.
Information theory provides a selection of concepts to quantify both
the information gain and the state disturbance. The choice of measures
for these two effects is usually dictated by the relevance to a specific
application. In most cases, however, derivation of the actual balance
represents a highly nontrivial task, especially if one is tempted
to resign from numerical means. The purpose of this Letter is to 
present a formulation of the information gain versus state disturbance
trade-off which is completely solvable using elementary analytical
techniques. This formulation is motivated by recent works on quantum
state estimation
\cite{Estimation}, where the information obtained from the operation
is converted into an estimate for the initial state of the system.

The problem considered in this Letter can be
formulated as follows. Suppose we are given a single $d$-level particle
in a completely unknown pure state $|\psi\rangle$. We want to make
a guess about the quantum state of this particle, but at the same time
we would like to alter the state as little as possible. One can
associate two fidelities with such a procedure. The first one, which
we will denote by $F$, describes how much the state after the operation
resembles the original one. The second fidelity, denoted by $G$,
characterizes the average quality of our guess. It is natural to expect
that these two quantities cannot take simultaneously too large values.
What is the actual quantitative bound between them?

Two extreme
cases are well known: if nothing is done to the particle we have $F=1$,
but then our guess about the state of the particle has to be random, which
yields $G=1/d$. On the other hand, the optimal estimation strategy for a
single copy
\cite{DDimensions}
yields $G=2/(d+1)$, but then the particle
after the operation cannot provide any more information on the initial
state; thus also $F=2/(d+1)$. I prove here that quantum
mechanics imposes a general constraint
between $F$ and $G$ in the form of the following inequality:
\begin{equation}
\label{Eq:TRADEOFF}
\sqrt{F-\frac{1}{d+1}} \le \sqrt{G - \frac{1}{d+1}}
+ \sqrt{(d-1)\left( \frac{2}{d+1} - G \right)}.
\end{equation}
I also show that this inequality cannot be further
improved, i.e.\ there exist quantum operations saturating the
equality sign.

The most general strategy that can be applied to
the particle has the form of a trace-preserving
operation described by a set of operators $\hat{A}_r$, where
$r=1,\ldots,N$. These operators satisfy the completeness relation:
\begin{equation}
\label{Eq:TracePreserving}
\sum_{r=1}^{N} \hat{A}_r^\dagger \hat{A}_r = \hat{\openone}.
\end{equation}
The classical information gained from this operation is given by
the index $r$, which is subsequently used to estimate the initial state
of the particle.
The outcome $r$ of the operation performed on a state
$|\psi\rangle$ is obtained with the probability
$\langle \psi | \hat{A}_r^\dagger \hat{A}_r | \psi \rangle$. This
corresponds to the following conditional transformation of the quantum
state:
\begin{equation}
|\psi\rangle \rightarrow \frac{\hat{A}_r |\psi\rangle}{
\sqrt{\langle \psi | \hat{A}_r^\dagger \hat{A}_r | \psi \rangle}}.
\end{equation}
We shall measure the resemblance of the transformed state to the original
one using the squared modulus of the scalar product, equal
$|\langle \psi | \hat{A}_r | \psi \rangle |^2/\langle \psi |
\hat{A}_r^\dagger \hat{A}_r | \psi \rangle$. Summation of this
expression over $r$ with the weights $\langle \psi |
\hat{A}_r^\dagger \hat{A}_r | \psi \rangle$, and integration
over all possible input states $|\psi\rangle$, yields the complete
expression for the mean operation fidelity $F$:
\begin{equation}
\label{Eq:Fdef}
F = \int \text{d}\psi  \sum_{r=1}^{N} | \langle \psi |
\hat{A}_r | \psi \rangle |^2.
\end{equation}
Here the integral $\int\text{d}\psi$ over the space of pure states
is performed using the canonical measure invariant with respect
to the group unitary transformations on the state vectors of the
particle.

Given the outcome $r$ of the operation, we can make a guess $|\psi_r\rangle$
what the state originally was. The quality of this guess, assuming that
the initial state was $|\psi\rangle$, can be quantified with the
help of the
overlap $|\langle \psi_r | \psi \rangle|^2$. The mean estimation fidelity
$G$ is given by the average of this expression over all outcomes $r$ with
the probability distribution $\langle \psi |
\hat{A}_r^\dagger \hat{A}_r | \psi \rangle$, and by integration over
states $|\psi\rangle$:
\begin{equation}
\label{Eq:Gdef}
G = \int \text{d} \psi \sum_{r=1}^{N}
\langle \psi | \hat{A}_r^\dagger \hat{A}_r | \psi \rangle
\, | \langle \psi_r 
| \psi \rangle|^2.
\end{equation}

We will start derivation of the trade-off between the fidelities
$F$ and $G$ by evaluating the integrals over $|\psi\rangle$.
For this purpose, let us introduce
in Eq.~(\ref{Eq:Fdef}) two decompositions of unity
in a certain orthonormal basis $|i\rangle$:
\begin{eqnarray}
F & = & \sum_{r=1}^{N} \sum_{i,j=0}^{d-1} \langle \psi | i \rangle
\langle i | \hat{A}_{r}^\dagger | \psi \rangle \langle \psi | \hat{A}_r
| j \rangle \langle j | \psi \rangle
\nonumber \\
& = & \sum_{r=1}^{N} \sum_{i,j=0}^{d-1} \langle i | 
\hat{A}_{r}^\dagger
\hat{M}_{ij}
\hat{A}_r
| j \rangle
\end{eqnarray}
where by $\hat{M}_{ij}$ we have denoted the following
integrals of projectors on the states $|\psi\rangle\langle\psi|$:
\begin{equation}
\hat{M}_{ij} = \int \text{d} \psi \,
\langle \psi | i \rangle \langle j | \psi \rangle
\, |\psi\rangle \langle
\psi|
=
\frac{1}{d(d+1)} (\delta_{ij} \hat{\openone} + | i \rangle \langle j |).
\end{equation}
The second explicit form of the operators $\hat{M}_{ij}$ has
been derived in Ref.~\cite{BanaXXX00}. This formula allows us to simplify
the expression for the mean operation fidelity $F$ to the form:
\begin{eqnarray}
F & = &
\frac{1}{d(d+1)} \left( \sum_{i=0}^{d-1} \sum_{r=1}^{N}
\langle i | \hat{A}_r^\dagger \hat{A}_r |i \rangle
+ \sum_{r=1}^{N} \left| \sum_{i=0}^{d-1} \langle i |
\hat{A}_r | i \rangle \right|^2 \right)
\nonumber \\
\label{Eq:FTrace}
& = &
\frac{1}{d(d+1)} \left(d + \sum_{r=1}^{N} | \text{Tr} \hat{A}_r |^2
\right)
\end{eqnarray}

Let us now consider the estimation fidelity $G$. The guess
$|\psi_r\rangle$ can be represented as a result of a certain unitary
transformation $\hat{U}_r$ acting on a reference state, which we will
take for concreteness to be $|0\rangle$:
\begin{equation}
|\psi_r\rangle = \hat{U}_r |0\rangle
\end{equation}
Using this representation, and 
changing the integration measure in Eq.~(\ref{Eq:Gdef})
according to $|\psi\rangle \rightarrow \hat{U}_r |\psi\rangle$,
we can evaluate the integral over $|\psi\rangle$:
\begin{eqnarray}
G & = & \sum_{r=1}^{N} \int \text{d} \psi | \langle 0 | \psi \rangle |^2
\,
\langle \psi | \hat{U}_r^\dagger \hat{A}_r^\dagger \hat{A}_r
\hat{U}_r | \psi \rangle
\nonumber \\
& = & \sum_{r=1}^{N}
\text{Tr} ( \hat{U}_r^\dagger \hat{A}_r^\dagger \hat{A}_r
\hat{U}_r  \hat{M}_{00} )
\end{eqnarray}
Inserting the explicit form of the operator $\hat{M}_{00} =
(\hat{\openone} + |0\rangle\langle 0|)/[d(d+1)]$ yields:
\begin{eqnarray}
G & = &
\frac{1}{d(d+1)}
\left(
\sum_{r=1}^{N} \text{Tr} ( \hat{U}_r^\dagger \hat{A}_r^\dagger \hat{A}_r
\hat{U}_r)
+
\sum_{r=1}^{N} \langle 0 | \hat{U}_r^\dagger \hat{A}_r^\dagger \hat{A}_r
\hat{U}_r | 0 \rangle
\right)
\nonumber \\
& = &
\frac{1}{d(d+1)} \left( d + \sum_{r=1}^{N} \langle \psi_r |
\hat{A}_r^\dagger \hat{A}_r
| \psi_r \rangle \right)
\end{eqnarray}
This expression provides directly a recipe for optimal assignment
of guesses $|\psi_r\rangle$ to outcomes of the operation:
each of the components
$\langle \psi_r |
\hat{A}_r^\dagger \hat{A}_r
| \psi_r \rangle $
in the sum over $r$ is maximized if $
| \psi_r \rangle $ is the eigenvector of $\hat{A}_r^\dagger \hat{A}_r$
corresponding to its maximum eigenvalue. 
Consequently, the maximum value of the mean
estimation fidelity $G$ for a given operation $\{\hat{A}_r\}$ can
be written as:
\begin{equation}
G = \frac{1}{d(d+1)} \left( d + \sum_{r=1}^{N} \| \hat{A}_r \|^2 \right)
\end{equation}
where the operator norm is defined in the standard way:
\begin{equation}
\| \hat{A}_r \| = \sup_{\langle \varphi | \varphi \rangle = 1}
\sqrt{\langle \varphi |  \hat{A}^\dagger_r \hat{A}_r | \varphi \rangle}.
\end{equation}

In order to relate the fidelities $F$ and $G$ to each other,
let us consider a polar decomposition of the operators $\hat{A}_r$:
\begin{equation}
\hat{A}_r = \hat{V}_r \hat{D}_r \hat{W}_r
\end{equation}
where $\hat{V}_r$ and $\hat{W}_r$ are unitary, and $\hat{D}_r$ is
a semi-positive definite diagonal matrix:
\begin{equation}
\hat{D}_r = \sum_{i=0}^{d-1} \lambda^r_i |i \rangle \langle i |,
\end{equation}
with the diagonal
elements put in a decreasing order: $\lambda^{r}_{0} \ge \ldots
\ge \lambda^{r}_{d-1} \ge 0$. We will first 
show that only the diagonal matrices $\hat{D}_r$ are relevant 
to the trade-off. Indeed,
the modulus of the
trace of the matrix $\hat{A}_{r}$ appearing in
Eq.~(\ref{Eq:FTrace}) is bounded by:
\begin{eqnarray}
|\text{Tr}\hat{A}_r| & = & \left| \sum_{i=0}^{d-1} \langle i
| \hat{W}_r \hat{V}_r
\hat{D}_r | i \rangle  \right| 
\nonumber \\
& \le & \sum_{i=0}^{d-1}
\lambda^{r}_{i}
| \langle i | \hat{W}_r \hat{V}_r | i \rangle |
\le
\sum_{i=0}^{d-1} \lambda^{r}_{i}
\end{eqnarray}
and moreover any quantum operation can be easily modified in such
a way that the equality sign is reached. What needs
to be done, is
to follow the operation $\{\hat{A}_r\}$ with
an extra unitary transformation $\hat{W}_r^\dagger
\hat{V}_r^\dagger$ depending on the outcome $r$. Let us note
that this corresponds to the modification of the operation
according to
$\hat{A}_r \rightarrow \hat{W}_r^\dagger \hat{V}^\dagger_r \hat{A}_r$,
which makes each element of the operation
a semi-positive hermitian operator.
As we are interested in the maximum value of $F$, we can further
assume with no loss of generality that:
\begin{equation}
\label{Eq:Flambda}
F  =  \frac{1}{d(d+1)} \left[ d + \sum_{r=1}^{N}
\left( \sum_{i=0}^{d-1} \lambda_{i}^{r} \right)^2 \right].
\end{equation}
The expression for the estimation fidelity written in terms of
$\lambda_i^r$ takes the form:
\begin{equation}
\label{Eq:Glambda}
G  =  \frac{1}{d(d+1)} \left( 
d + \sum_{r=1}^{N} (\lambda_0^r)^2 \right).
\end{equation}
In addition, the trace of the
completeness condition given in Eq.~(\ref{Eq:TracePreserving}) yields
the following constraint on $\lambda_{i}^{r}$:
\begin{equation}
\label{Eq:TracePresLambda}
\sum_{r=1}^{N} \sum_{i=0}^{d-1} ( \lambda_{i}^{r} )^2 = d.
\end{equation}
To complete the proof of the inequality (\ref{Eq:TRADEOFF}),
it is convenient to introduce vector notation. Let us define
$d$ real
vectors ${\bf v}_i = (\lambda^{1}_{i}, \ldots \lambda^{N}_{i})$,
where the index $i$ runs from $0$ to $d-1$.
Sums over $r$ appearing in Eqs.~(\ref{Eq:Flambda})
and (\ref{Eq:Glambda}) can be written as:
\begin{eqnarray}
\label{Eq:f}
f & = & \sum_{r=1}^{N}
\left( \sum_{i=0}^{d-1} \lambda_{i}^{r} \right)^2
= \sum_{i,j=0}^{d-1} {\bf v}_i \cdot {\bf v}_j
\\
g & = & 
\sum_{r=1}^{N} (\lambda_0^r)^2 = | {\bf v}_0 |^2
\end{eqnarray}
where the dot denotes the scalar product, and $|\cdot|$ is the
standard quadratic norm.
The completeness condition (\ref{Eq:TracePresLambda})
for the operation $\{\hat{A}_r\}$ written in the vector
notation takes the form
\begin{equation}
\label{Eq:d}
\sum_{i=0}^{d-1} | {\bf v}_i |^2 = d.
\end{equation}
Let us now suppose that the vector ${\bf v}_0$ is fixed. The estimation
fidelity is then given by $G=(d+|{\bf v}_0|^2)/[d(d+1)]$.
What is the maximum
operation fidelity $F$ that can be achieved with this constraint? 
The answer to this question is provided by an
application of the Schwarz inequality to Eq.~(\ref{Eq:f}):
\begin{equation}
\label{Eq:fle}
f 
\le \sum_{i,j=0}^{d-1} |{\bf v}_i| |{\bf v}_j| =
\left( \sum_{i=0}^{d-1} |{\bf v}_i| \right)^2
= \left(\sqrt{g} + \sum_{i=1}^{d-1} |{\bf v}_i| \right)^2
\end{equation}
We have excluded here from the sum over $i$ the norm of the vector
${\bf v}_0$ which is fixed and equal to
$\sqrt{g}$. The sum of the norms of the remaining
vectors can be estimated using the inequality between the arithmetic
and quadratic means:
\begin{equation}
\label{Eq:ArithQuadMeans}
\frac{1}{d-1}\sum_{i=1}^{d-1} 
| {\bf v}_i|
\le \sqrt{ \frac{1}{d-1}
\sum_{i=1}^{d-1} |{\bf v}_i|^2 } =
\sqrt{\frac{d-g}{d-1}},
\end{equation}
where we have evaluated the sum $\sum_{i=1}^{d-1} |{\bf v}_i|^2$
using Eq.~(\ref{Eq:d}). Inserting this bound into Eq.~(\ref{Eq:fle})
we finally obtain the inequality
\begin{equation}
f \le \left(\sqrt{g} + \sqrt{(d-1)(d-g)}\right)^2
\end{equation}
which expressed in terms of the fidelities
$F$ and $G$ takes the form of Eq.~(\ref{Eq:TRADEOFF}).

The necessary and sufficient conditions for a quantum operation
to reach the equality sign can be most easily formulated in the vector
notation. The Schwarz inequality (\ref{Eq:fle}) becomes equality
if all the vectors ${\bf v}_0, \ldots , {\bf v}_{d-1}$ are collinear.
Furthermore, equation sign in Eq.~(\ref{Eq:ArithQuadMeans}) holds
if and only if $|{\bf v}_1| = \ldots = |{\bf v}_{d-1}|$.
It is straightforward to see that an exemplary operation
satisfying these conditions for a given estimation fidelity
$G=(1+g/d)/(d+1)$ is defined by:
\begin{equation}
\hat{A}_r = \sqrt{\frac{g}{d}} |r-1\rangle \langle r-1|
+ \sqrt{\frac{d-g}{d(d-1)}}(\hat{\openone} - |r-1\rangle \langle r-1|)
\end{equation}
where the index $r$ runs from $1$ to $d$, and the projectors
$|r-1\rangle\langle r-1|$ are constructed using any orthonormal basis.
This confirms the inequality (\ref{Eq:TRADEOFF}) is indeed a tight one
and cannot be further improved.

A simple transformation of Eq.~(\ref{Eq:TRADEOFF}) shows that
the quantum mechanically allowed region for the fidelities
$F$ and $G$ is bounded by a quadratic curve, which turns out
to be a fragment of an ellipse given by the equation:
\begin{equation}
\begin{array}[b]{lcl}
(F-F_0)^2 + d^2 (G-G_0)^2 & & \\
+ 2 (d-2) (F- F_0)(G-G_0) & = &
\displaystyle
\frac{d-1}{(d+1)^2}
\end{array}
\end{equation}
with $F_0 = (d+2)/(2d+2)$ and $G_0=3/(2d+2)$. The shape of the 
region for several values of $d$ is depicted in Fig.~1.

The balance between the operation and estimation fidelities derived
in this Letter has interesting consequences in quantum
teleportation based on
nonmaximally entangled states. If two parties share a pure bipartite
state of the Schmidt form $|\text{tele}\rangle
= \sum_{k=0}^{d-1} \mu_k |k\rangle \otimes |k\rangle$,
then the maximum teleportation fidelity attainable using
this state is given by \cite{VidaJonaXXX99,BanaXXX00}:
\begin{equation}
F_{\text{tele}} = \frac{ 1 + \left(
\sum_{k=0}^{d-1} \mu_k \right)^2}{d+1}.
\end{equation}
Furthermore, for a nonmaximally entangled state the measurement performed
during the teleportation protocol reveals some information on the teleported
state. This information can be converted into an estimate
for the initial state, whose maximum average fidelity has been shown
to equal \cite{BanaXXX00}:
\begin{equation}
G_{\text{tele}} = \frac{1 + \mu_0^2}{d+1}
\end{equation}
where $\mu_0$ denotes the largest Schmidt coefficient for the state
$|\text{tele}\rangle$. As the procedure of teleportation can be viewed
as a special case of a quantum operation \cite{NielCavePRA97},
the bound (\ref{Eq:TRADEOFF}) applies as well to the pair of fidelities
$F_{\text{tele}}$ and $G_{\text{tele}}$. Consequently, for a given
teleportation fidelity $F_{\text{tele}}$, the maximum value of the
estimation fidelity is achieved for the state $|\text{tele}\rangle$
satisfying the condition $\mu_1 = \ldots = \mu_{d-1} = 
\sqrt{(1-\mu_0^2)/(d-1)}$. This condition defines a class of pure
bipartite states which are optimal from the point of view of
the trade-off between the teleportation fidelity and the estimation
fidelity.

In conclusion, I have obtained a tight bound for the fidelities
describing the quality of estimating the state
of a single copy of a $d$-level particle, and the degree the initial
state has to be changed during this operation. This result seems
to be one of very few cases, when the trade-off between the
information gain and the state disturbance can be derived in a closed
analytical form.

I would like to acknowledge useful discussions with J. H. Eberly,
C. A. Fuchs, N. L\"{u}tkenhaus, V. Vedral, and I. A. Walmsley.
This research was partially supported by ARO--administered
MURI grant No.\ DAAG-19-99-1-0125, NSF grant PHY-9415583, and
KBN grant 2~P03B~089~16.

\begin{figure}
\centerline{\epsfig{file=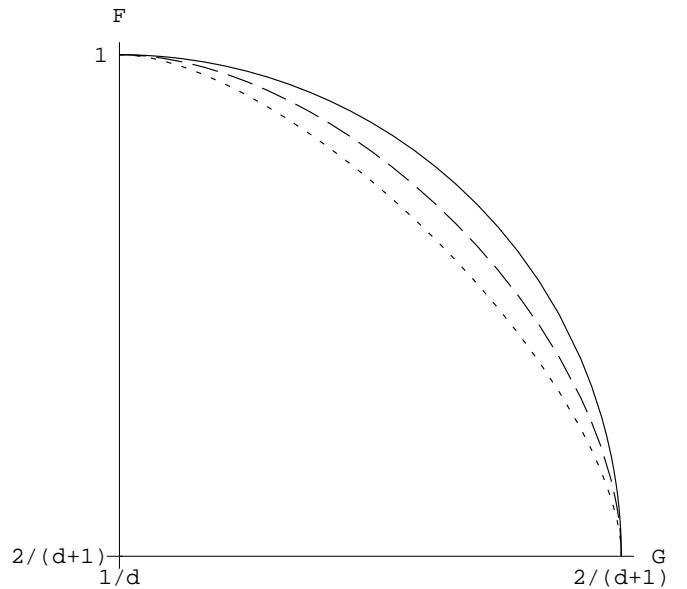,width=3.5in}}

\vspace{5mm}

\caption{Rescaled bound for the operation fidelity $F$ versus
the estimation fidelity $G$, plotted for $d=2$ (solid line), $d=4$
(dashed line), and $d=8$ (dotted line).}
\end{figure}

\end{document}